\documentclass[]{article}
\usepackage{graphicx}

\usepackage{amssymb}
\usepackage{amsthm}
\usepackage{xcolor}
\usepackage{authblk}

\usepackage{lineno}

\begin{document}




\title{{\tt PyPVRoof}: a Python package for extracting the characteristics of rooftop PV installations using remote sensing data}

\date{}

\author[1]{Yann Tremenbert\footnote{Work done while in internship at RTE}}
\author[1,2]{Gabriel Kasmi}
\author[2]{Laurent Dubus}
\author[1]{Yves-Marie Saint-Drenan}
\author[1]{Philippe Blanc}

\affil[1]{MINES Paris, Université PSL, Centre Observation Impacts Energie (O.I.E.), 06904 Sophia Antipolis, France}
\affil[2]{RTE - Réseau de transport d'électricité, 92073 Paris La Défense, France}

\maketitle

\begin{abstract}
    Photovoltaic (PV) energy grows at an unprecedented pace, which makes it difficult to maintain up-to-date and accurate PV registries, which are critical for many applications such as PV power generation estimation. This lack of qualitative data is especially true in the case of rooftop PV installations. As a result, extensive efforts are put into the constitution of PV inventories. However, although valuable, these registries cannot be directly used for monitoring the deployment of PV or estimating the PV power generation, as these tasks usually require PV systems {\it characteristics}. To seamlessly extract these characteristics from the global inventories, we introduce {\tt PyPVRoof}. {\tt PyPVRoof} is a Python package to extract essential PV installation characteristics. These characteristics are tilt angle, azimuth, surface, localization, and installed capacity. {\tt PyPVRoof} is designed to cover all use cases regarding data availability and user needs and is based on a benchmark of the best existing methods. Data for replicating our accuracy benchmarks are available on our Zenodo repository \cite{tremenbert2023pypvroof}, and the package code is accessible at this URL: {\tt https://github.com/gabrielkasmi/pypvroof}.
\end{abstract}

\section{Introduction}


Photovoltaic (PV) installed capacity grows quickly \cite{rte2021bilan,iea2021solar} as it is key to decarbonizing energy systems \cite{haegel2017terawatt}. Keeping track of the deployment and characteristics of the PV installed capacity can be difficult, and public authorities and industrial stakeholders often lack precise knowledge regarding the PV fleet \cite{kasmi2023crowdsourced}. This concern is especially true in the case of rooftop PV, for which no centralized and disaggregated registry are generally available \cite{kasmi2022towards}. The lack of reliable information on the PV installed capacity can yield unreliable rooftop PV power generation estimates. For instance, for transmission system operators (TSOs), the lack of reliable rooftop PV measurements increases the flexibility needs, i.e., the ability of the grid to compensate for load or supply variability \cite{kazmi2022good,saint2016analysis,saint2019bayesian,huber2014integration}. Therefore, there is a growing need for reliable and more accurate rooftop PV mapping and characterization, i.e., estimation of the technical characteristics of the installation.

Numerous efforts were carried out to acquire information regarding PV installations in general and distributed PV in particular. For instance, Dunnett et al. \cite{dunnett2020harmonised} proposed a harmonized dataset using publicly available data from OpenStreetMap. Stowell et al. \cite{stowell2020harmonised} leveraged crowdsourcing to map approximately 86\% of the United Kingdom's distributed PV installed capacity. In their approach, users were asked to delineate PV panels in their neighborhood. Alternatively, several works leveraged deep learning and overhead imagery to map PV installations quickly \cite{malof2019mapping,yu2018deepsolar,mayer2020deepsolar,hu2019you}. Usually, a model is trained on a training dataset (e.g., \cite{bradbury2016distributed,kasmi2023crowdsourced,khomiakov2022solardk}) and then deployed on a larger area. Overhead imagery comes as large tiles, so they are cut into small patches and passed into the trained model, which will return the probability that each image pixel depicts a PV panel. One finally transforms the so-called resulting segmentation map into a set of geolocalized polygons of the same format as the crowdsourced works previously mentioned.

Although valuable, PV polygons are insufficient for many applications (e.g., PV power generation estimation \cite{saint2017probabilistic}. In order to provide installation characteristics, several methods have been proposed. Edun et al. \cite{edun2021unsupervised} leveraged the Hough transform for azimuth estimation. Mayer et al. \cite{mayer20223d} used heuristics and LiDAR surface models to create PV registries containing tilt, azimuth, and installed capacity. So et al. \cite{so2017estimating} showed that installed capacity could be derived from the surface using a linear model. Recently, Perry et al. \cite{perry2023panel} introduced a python package to extract information such as the mounting configuration of PV panels. The main limitation of these works is that they have different data requirements, which limits their reproducibility.

To standardize the estimation of rooftop PV systems characteristics, we introduce {\tt PyPVRoof}. This Python package is designed to estimate the technical characteristics (tilt and azimuth angles, installed capacity, localization, surface) from its geolocalized polygon and additional input data, i.e., preexisting PV registries and/or digital surface models (DSMs). {\tt PyPVRoof} combines characteristics extraction methods chosen after a careful benchmark on the training dataset BDAPPV \cite{kasmi2023crowdsourced}, which contains PV segmentation masks and installation characteristics. We designed this package to work under the most common scenarios of data availability.

To the best of our knowledge, {\tt PyPVRoof} is the first attempt towards a standardized package for the extraction of PV systems characteristics. It works with polygons, the usual way of encoding PV systems localization, and handles different additional data sources to work even if the user does not have digital surface models (DSMs) or existing data sources. 

The remainder is organized as follows: in section \ref{sec:literature}, we review existing works focusing on PV mapping and characteristics extraction. In section \ref{sec:testbench} we present how our package works, its methods and metrics used for evaluation. In section \ref{sec:data}, we review the data that we use for our benchmark and that the package requires to work. In section \ref{sec:results}, we present our benchmark results and discuss the final choice of methods included in {\tt PyPVRoof}. Section \ref{sec:conclusion} concludes.

\section{Related works}\label{sec:literature}

\subsection{Inventories of PV systems} 

Keeping track of the deployment of PV installation is challenging as it is growing at an unprecedented pace. It is especially true in the case of distributed PV since the latter is aggregated at the city or census scale in current available public registries, with no information regarding the characteristics of the underlying installations \cite{de2020monitoring,kasmi2022towards}. As a response, numerous initiatives emerged to map PV installations, ranging from power plants \cite{kruitwagen2021global,dunnett2020harmonised,hou2019solarnet,plakman2022solar} to rooftop PV installations \cite{malof2019mapping,yu2018deepsolar,mayer2020deepsolar,hu2019you,kasmi2022towards,mayer20223d,stowell2020harmonised}. When released, these registries take the form of geolocalized polygons, usually in the {\tt .geojson} format. These polygons can be accessed through online platforms such as OpenInfraMap \cite{openinframap}. These inventories come from automatic detection from overhead imagery (\cite{mayer2020deepsolar,kasmi2022towards}) or directly through crowdsourcing efforts (\cite{stowell2020harmonised}).

\subsection{PV systems characteristics extraction}

In addition to detecting and mapping PV installations, extracting the characteristics has greatly interested researchers and public authorities. Indeed, the installed capacity, expressed in kWp, is the primary indicator to keep track of the deployment of PV \cite{chen2023remote}. As a response, several methods were proposed to extract PV characteristics in addition to geolocalized polygons. So et al. \cite{so2017estimating} estimated the installed capacity from the polygon's surface using a linear regression model. Rausch et al. \cite{rausch2020enriched} and Mayer et al. \cite{mayer20223d} further refined this approach to propose a method for automatically constructing a detailed registry using overhead imagery and 3D data. More recently, Kasmi et al. \cite{kasmi2022towards} proposed a method for mapping and characterizing PV installations without needing surface models. A few works also proposed methods to extract the azimuth angle of the installations (\cite{edun2021unsupervised}) or the tilt and azimuth category (i.e., angle and azimuth ranges, \cite{memari2022deep}). More recently \cite{perry2023panel} proposed their package for extracting PV systems characteristics using deep learning and satellite imagery. Their package extracts the azimuth angle using the method of Edun et al. \cite{edun2021unsupervised} and the configuration type (mounted, rooftop, tracker). However as their approach only relies on satellite images, they do not extract other characteristics, in particular the tilt angle of the installation, leading to an incomplete characterization of the PV installations. We show that a comprehensive characterization requires additional data sources (i.e., auxiliary data or digital surface models), in particular for the tilt angle and installed capacity estimation.

\section{Methods}\label{sec:testbench}

\subsection{{\tt PyPVRoof}: a Python package for extracting PV installations characteristics from geolocalized polygons}

\subsubsection{Overview}

\begin{figure}[h]
    \centering
    \includegraphics[width = \textwidth]{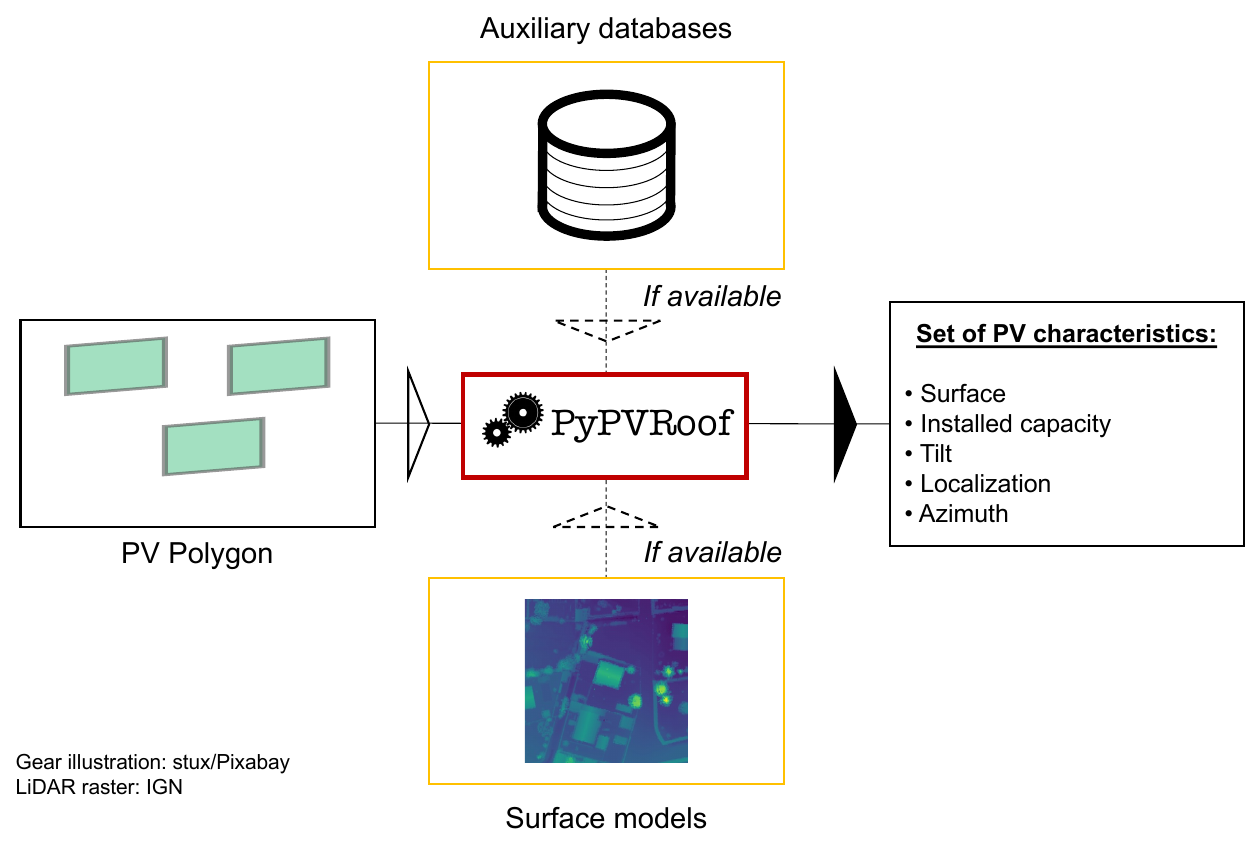}
    \caption{Flowchart of the proposed method to extract installations' characteristics}
    \label{fig:flowchart}
\end{figure}

Figure \ref{fig:flowchart} summarizes the workflow of {\tt PyPVRoof}. {\tt PyPVRoof} extracts PV characteristics from geolocalized polygons. It accommodates additional data sources, such as preexisting registries (i.e., auxiliary data) or digital surface models (DSM), depending on their availability for the user. The list of characteristics that we extract is the following:

\begin{itemize}
    \item Localization (latitude and longitude)
    \item Tilt angle (in degrees)
    \item Azimuth angle (in degrees, relative to North)
    \item Surface (in m²). Estimating the surface requires knowing the tilt, as only the {\it projected} surface is derived from the input polygon.
    \item Installed capacity (in kWp). The surface is needed to estimate the installed capacity as its first-order approximation is the surface multiplied by an efficiency factor \cite{so2017estimating}.
\end{itemize}

This set of characteristics covers various use cases, ranging from PV systems inventories to PV power forecasting. It is sufficient to describe an installation using physical models \cite{killinger2018search}. To extract all characteristics mentioned above or only some of them, the user only has to specify a method (provided that the data requirements are satisfied) and pass the polygon as input. Further details and tutorials are accessible on the public repository, accessible at this URL: 
{\tt https://github.com/gabrielkasmi/pypvroof}.

\subsubsection{{\tt PyPVRoof} combines best-in-their-class methods}

{\tt PyPVRoof} combines methods for characteristics extraction based on a review of existing works in the field. We reviewed the main methods used, evaluated their data requirements, and benchmarked them to keep the best-performing methods. Further details about our benchmark results are provided in section \ref{sec:results}. These methods were chosen based on accuracy, simplicity, and efficiency. In particular, we retained the most simple between two equally performing methods (e.g., the look-up table over the random forest). Finally, we restricted ourselves to methods that require as few additional inputs as possible. These methods reflect the current state-of-the-art for characteristics extraction. 

\subsubsection{An approach that adapts to the available data}\label{sec:pypvroof-methods}

As characteristics extraction is usually part of a larger pipeline (e.g., remote PV mapping such as in Mayer et al. \cite{mayer20223d}, or Kasmi et al. \cite{kasmi2022towards}), resulting in multiple use cases depending on the approach proposed and the data available to the authors. Some models only rely on overhead imagery and only focus on surface estimation (\cite{yu2018deepsolar}), while other approaches are more comprehensive but also require more data (\cite{mayer20223d}). {\tt PyPVRoof} was developed to standardize existing approaches in both inputs and outputs. We also accommodate that additional data (e.g., auxiliary registries and/or surface models) can be considered during the process. Auxiliary data correspond to a registry containing PV characteristics, such as the characteristics file of \cite{kasmi2023crowdsourced}. The primary purpose of the auxiliary data file is to provide a first guess on the tilt angle and the panel efficiency value. Therefore, it can only be a sample of PV characteristics rather than a comprehensive registry. Surface models, i.e., digital surface models (DSM) correspond to a type of geographical information system (GIS) that delivers information about the height of objects (natural and built) on the ground. These models are often used to infer the tilt angle of buildings \cite{de2020quick,martin2020multi} or PV panels \cite{mayer20223d}.

\begin{figure}[h]
    \centering
    \includegraphics[width=\textwidth]{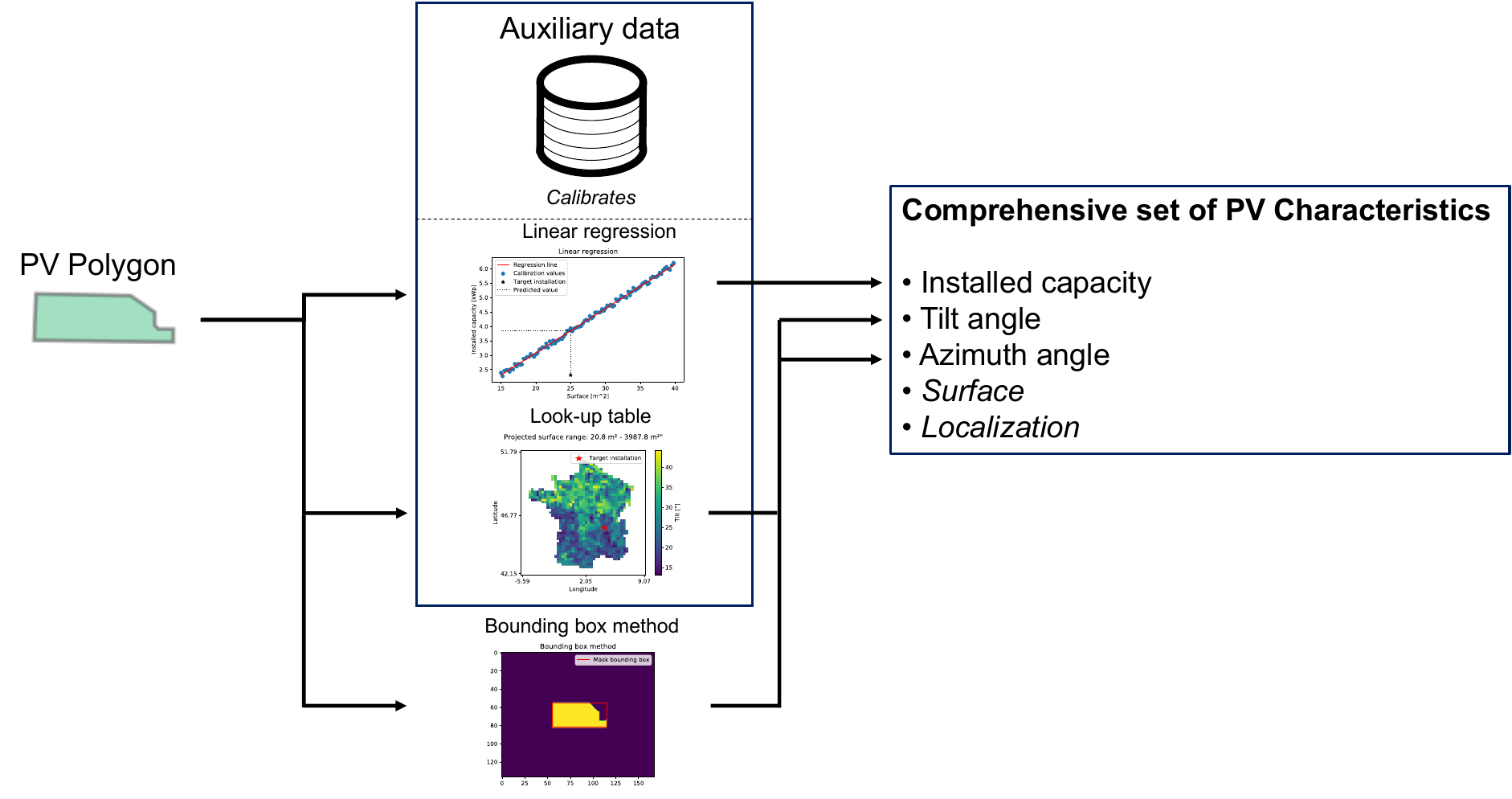}
    \caption{{\tt PyPVRoof} flowchart if only auxiliary data is available}
    \label{fig:case-1}
\end{figure}

\paragraph{Use-case 1: auxiliary data} Figure \ref{fig:case-1} presents the flowchart and the associated methods to extract PV characteristics if the user has only access to auxiliary data. In this case, {\tt PyPVRoof} leverages this data to calibrate the panel efficiency module coefficient to correlate the installation's surface with an installed capacity. A look-up table (LUT) is also computed from this input data for the tilt angle estimation. Finally, we apply a bounding box algorithm to estimate the azimuth angle.

\begin{figure}[h]
    \centering
    \includegraphics[width=\textwidth]{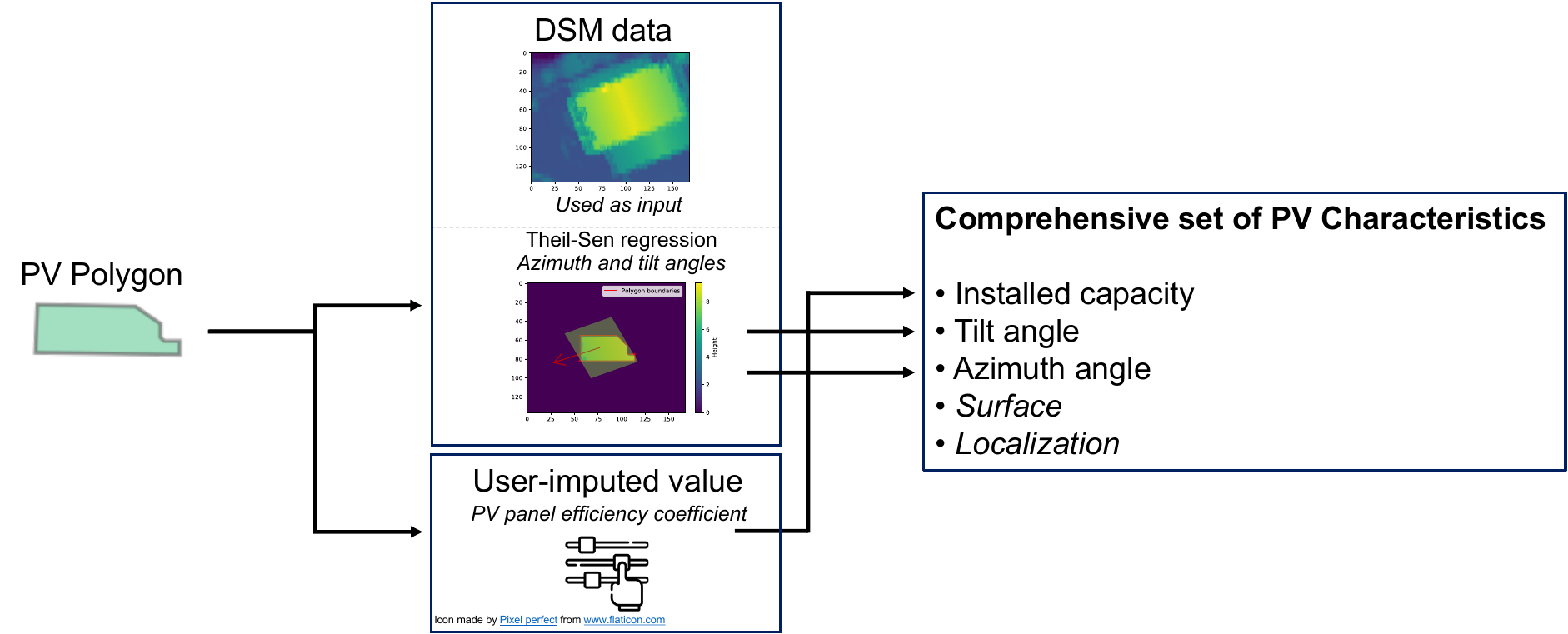}
    \caption{{\tt PyPVRoof} flowchart if only digital surface models (DSMs) are available}
    \label{fig:case-2}
\end{figure}

\paragraph{Use-case 2: DSM data} Figure \ref{fig:case-2} presents the flowchart and the associated methods to extract PV characteristics if the user can only access digital surface models. Using a Theil-Sen estimator, we leverage the DSM to estimate the tilt {\it and} azimuth angles. On the other hand, the panel efficiency coefficient is a parameter set by the user.

\begin{figure}[h]
    \centering
    \includegraphics[width=\textwidth]{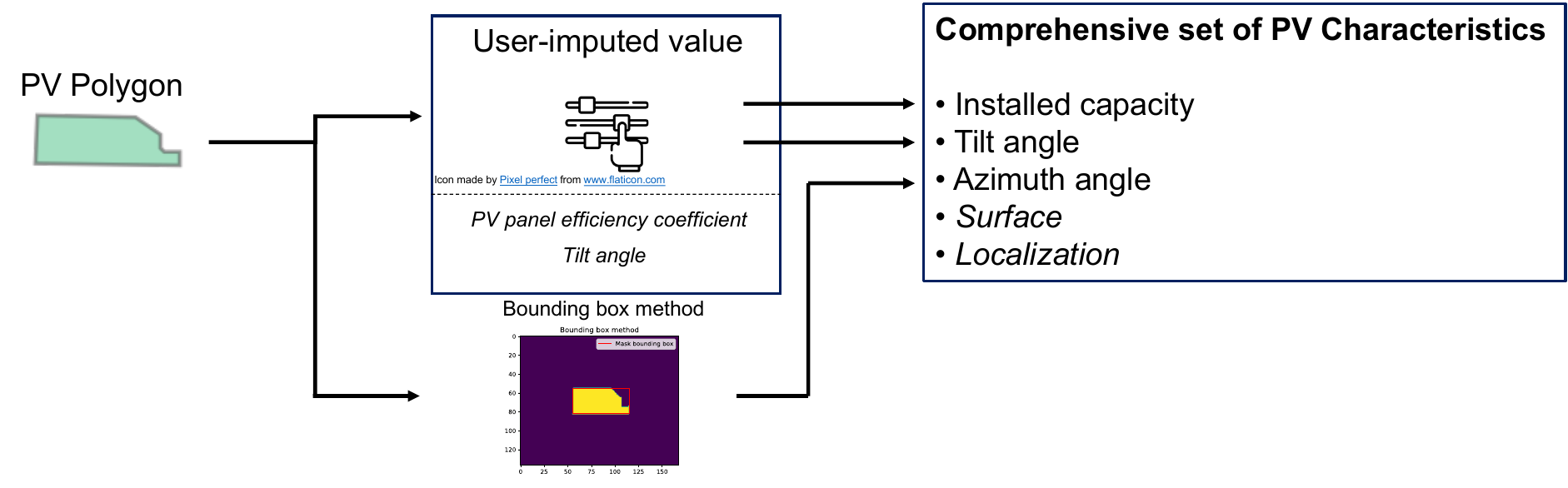}
    \caption{{\tt PyPVRoof} flowchart if no complementary data is available}
    \label{fig:case-3}
\end{figure}

\paragraph{Use-case 3: no data} Figure \ref{fig:case-3} presents the flowchart and the associated methods to extract PV characteristics if no auxiliary data is available. In this case, the user imputes the panel efficiency coefficient and an average tilt angle, but the azimuth angle is estimated using the bounding-box algorithm. 

In the project repository, we provide a tutorial that enables the user to try various methods to extract characteristics from polygons using different methods.

\subsection{Detailed presentation of the methods}\label{sec:methods}

In this section, we review the methods that we kept in {\tt PyPVRoof}. We refer the reader to the appendix \ref{sec:complementaty-methods} for a description of the remaining benchmarked methods.

\subsubsection{Direct computation}

Overhead imagery is usually orthorectified (i.e., with a uniform scale). One can only compute the {\it projected} surface from the polygon. The computation is straightforward, and the only requirement is considering the projection. Parhar et al. \cite{parhar2021hyperionsolarnet} describe the Mercator case. In practice, packages such as {\tt area} \cite{areapackage} estimate the surface of {\tt geojson} polygons, taking into account the deformation induced by the projection system.

Once the projected surface is known, one needs the tilt angle to compute the real surface. Denoting $S_{proj}$ the projected surface and $\theta$ the tilt angle of the installation (in degrees), the real surface is given by equation (\ref{eq:surface}).

\begin{equation}\label{eq:surface}
    S = S_{proj}/\cos\left(\theta \times \frac{\pi}{180}\right)
\end{equation}

\subsubsection{Constant parameters}

\paragraph{Constant tilt} Tilt is necessary to compute the real surface of the installation. When neither registries nor surface models are available, it is still possible to infer a tilt angle from the remaining data (i.e., the PV polygon). However, in practice, the optimal tilt angle of an installation is known. Typically, a tilt angle of around 30 degrees is optimal in most European countries. Regional models estimating the PV yield of solar plants consider this value by default \cite{jrc-tilts,saint2018approach}. In our case, we allow the user to input a default coefficient if necessary. This case can be seen as a worst-case situation if no surface models nor auxiliary data is available.

\paragraph{Constant efficiency} An efficiency factor relates the surface of a PV installation and its installed capacity. The PV panel efficiency increased due to the cell efficiency increase over the last couple of decades \cite{noauthor_best_nodate}. This efficiency is usually measured in kWp/m². The efficiency depends on many criteria (e.g., module technology of the panel, aging, manufacturer), which are not necessarily publicly available. However, average efficiencies can be used. For instance, \cite{rausch2020enriched,mayer20223d} used a value of 6 kWp/m²  as a reference value to estimate the installed capacity from the surface. As for the tilt angle, we allow the user to input this efficiency value.

\subsubsection{Theil-Sen estimation} 

The Theil-Sen estimator (TSE) initially proposed by Theil \cite{theil1992rank} and Sen \cite{sen1968estimates} is a robust regression method. It consists in considering the median of the slopes of all lines (or planes in higher dimensions) through pairs of points. This method is more robust to outliers than ordinary least squares. 

\begin{figure}[h]
    \centering
    \includegraphics[width=0.3\textwidth]{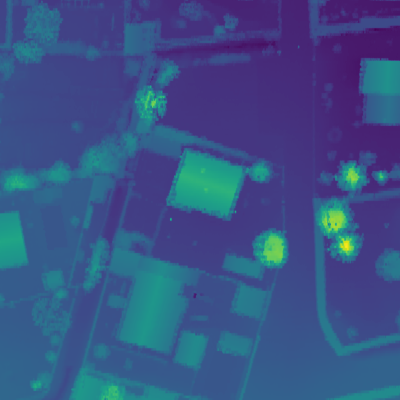}
    \caption{Example of DSM: the rasterization of the LiDAR from the IGN}
    \label{fig:lidar-example}
\end{figure}

We use this method to fit a plane $z(x,y) = ax + by + c$ parameterized by only three parameters, $a,b$ and $c$, to a set of points corresponding to altitudes. These altitudes come from the digital surface model (DSM) passed as input. Figure \ref{fig:lidar-example} depicts an example of LiDAR DSM provided by the IGN. Lighter areas correspond to higher altitudes.

\begin{figure}[h]
    \centering
    \includegraphics[width=0.6\textwidth]{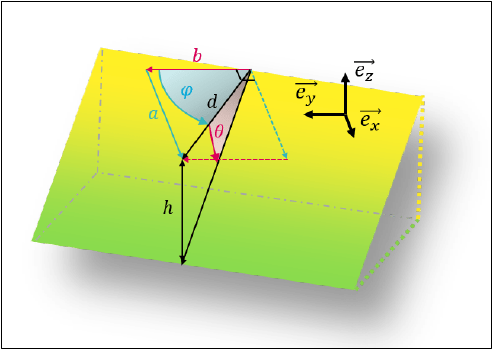}
    \caption{Theil-Sen method principle. The plane is deduced from the raster and is parameterized as $z(x,y) = ax + by + c$. $\varphi$ corresponds to the azimuth angle and $\theta$ to the tilt angle. $\overrightarrow{e_x}$, $\overrightarrow{e_y}$, $\overrightarrow{e_z}$ correspond to the canonical basis of $\mathbb{R}^3$.}
    \label{fig:theil-sen-principle}
\end{figure}

The direction of the gradient of the plane gives the azimuth angle $\varphi$. The slope value along this gradient corresponds to the tilt angle $\theta$. The gradient of the plane $\nabla z(x,y)$ is given in equation (\ref{eq:gradient-tse}):

\begin{equation}\label{eq:gradient-tse}
\nabla z (x,y) = \left(\frac{\partial z}{\partial  x}(x,y), \frac{\partial z}{\partial  y}(x,y)\right) = (a,b)
\end{equation}

and 

\begin{equation}\label{eq:formulas-tse} 
    \varphi = \arctan\left(\frac{a}{b}\right), \;
    \theta =  \arctan\left(\frac{h}{d}\right)
\end{equation}

where $h = a^2+b^2$ and $d = \sqrt{a^2+b^2}$. Figure \ref{fig:theil-sen-principle} depicts the principle of the Theil-Sen method to compute the tilt and azimuth angles.

\subsubsection{Look-up table}

If surface models are unavailable, we can still recover a tilt angle more accurately than with direct computation. To achieve this, we only need a sample of tilt angles for the desired area, e.g., a smaller PV database or a building database. We can then reflect the spatial variability of the tilt angle by computing an average tilt angle per grid point. The reference value associated with the installation corresponds to the average of the existing installations located in this grid point. The lookup table requires that the auxiliary data frame span the complete area or interest (e.g., a region or a country). 

We compute this so-called lookup table (LUT) only once, and the user can pass a precomputed LUT as input. Computation is done as follows: we first define the spatial extent by setting easternmost $E$, northernmost $N$, westernmost $W$, and southernmost $S$ boundaries. These boundaries are expressed in geographical coordinates. We then define a grid by dividing the numerical intervals defined by $E$ and $W$ and $S$ and $N$ respectively. We end up with $K$ longitude intervals and $L$ latitude intervals. Besides, we cluster the auxiliary data frame by (projected) surface category to define $T$ surface categories. After empirical investigations, defining intervals as quantiles yielded the best results.

We then aggregate all sample points $x_{1}^{k,l,t}, \dots x_{n}^{k,l,t}$ whose coordinates belong to the $k\times l-th$ grid point and projected surface that belong the $t-th$ category. We then compute the reference tilt angle for this $(k,l,t)-th$ box, denoting $\theta^{k,l,t}$ by averaging the tilt values of the $n$ sample points falling into this bin. If no sample is available, we do not input a value. 

\begin{figure}[h]
    \centering
    \includegraphics[width = 0.9\textwidth]{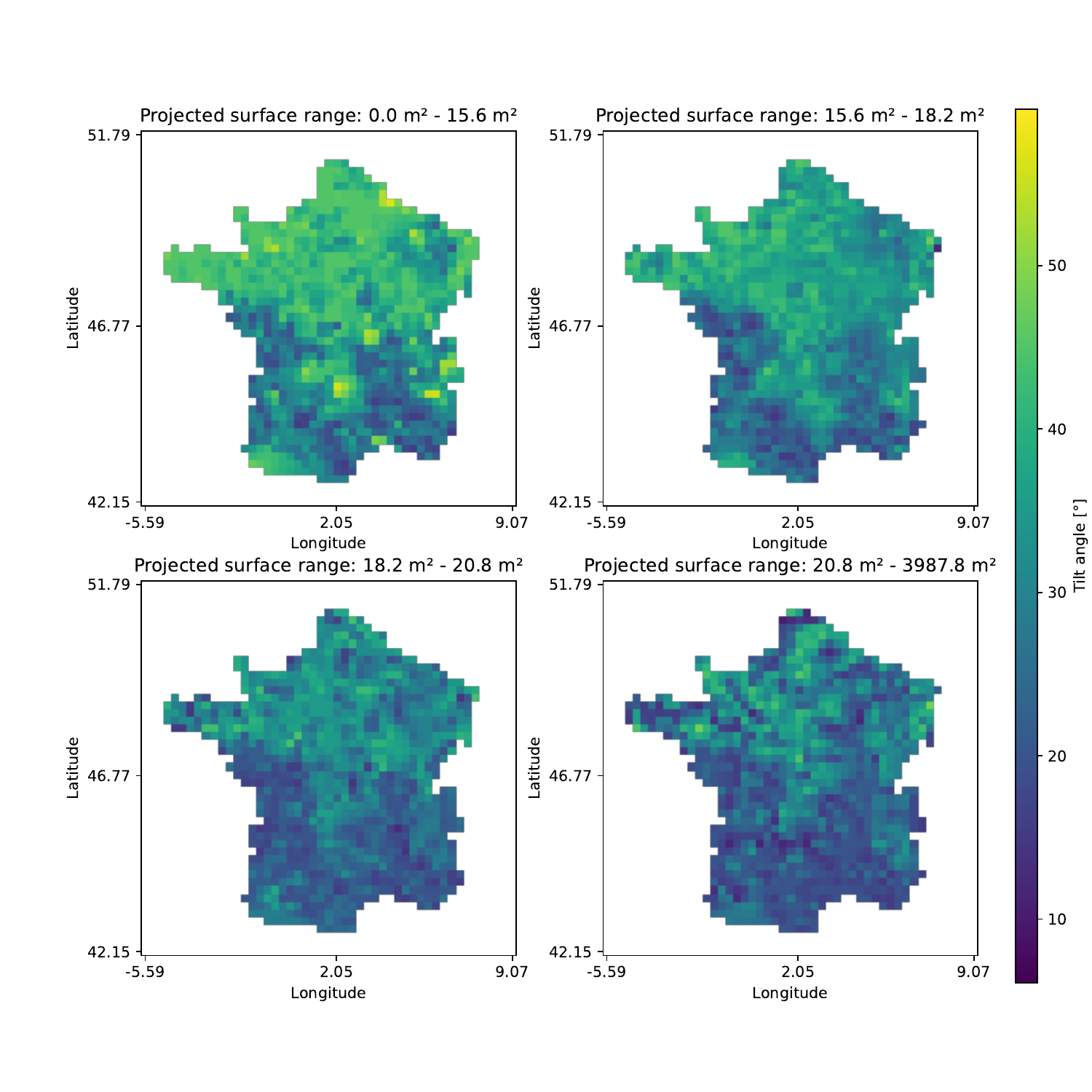}
    \caption{Lookup table for $50\times50$ grid-points and four surface categories computed for the PV mapping algorithm of Kasmi et al.\cite{kasmi2022towards}. Surface categories correspond to quartiles of the distribution of the surface in the auxiliary data.}
    \label{fig:lut-example}
\end{figure}

Once this step is finished, we end up with a subset of grid points for which no reference value is available. We estimate a value by interpolating a $\theta^{k,l,t}$ by interpolating the neighboring values. We do not interpolate across surface categories. Figure \ref{fig:lut-example} displays the LUT obtained for the PV mapping algorithm of Kasmi et al. (\cite{kasmi2022towards}) using this method. 

\subsubsection{Bounding-box}

The bounding box method only requires the polygon to compute the azimuth angle $\varphi$. The bounding-box method is an alternative when no surface models are accessible. We simplify the polygon's geometry by computing its bounding box. Then, we compute the azimuth angles associated with the "long" and "short" sides of the rectangle. We input as azimuth angle the angle corresponding to the longest side. We implicitly assume that the PV panel tends to be wider than high. The main limitation of this method is that it cannot distinguish between a panel facing eastwards or westwards, northwards or southwards. In the latter case, however, we can assume that the PV panel should not point northwards (at least in the Northern Hemisphere). If our bounding-box heuristic estimates that the polygon points between -45 and 45 degrees (0 being the reference for the North), we correct the estimation by applying a horizontal symmetry.

\subsubsection{Linear regressions}

So et al. \cite{so2017estimating} showed that it is possible to accurately estimate the installed capacity by fitting a linear regression between the surface and the installed capacity. We build on this method. The linear model is given by equation (\ref{eq:linear-reg-capacity}).

\begin{equation}\label{eq:linear-reg-capacity}
c = \gamma_0 + \gamma S    
\end{equation}
Where $S$ is the surface in m²  and $c$ is the capacity in kWp of the installation. As pointed out by So et al., \cite{so2017estimating}, $\gamma_0$ is a bias coefficient; in the true model, $\gamma_0$ should be equal to zero. In our case, we consider $\gamma_0=0$ and estimate $\gamma$ from BDAPPV. 

Efficiencies can differ depending on the PV installation's surface \cite{noauthor_best_nodate}. To accommodate this, we introduce another estimation for the installed capacity, namely the clustered linear regression. Clusters are defined depending on the surface of the installation. The goal is to reflect the different efficiencies while keeping the number of parameters as low as possible. This approach is inspired by the second model of  So et al. \cite{so2017estimating}, which estimated a panel-wise coefficient $\gamma$. Their approach, however, required additional unobservable information, such as the manufacturer's design.

\begin{figure}[h]
    \centering
    \includegraphics[width=\textwidth]{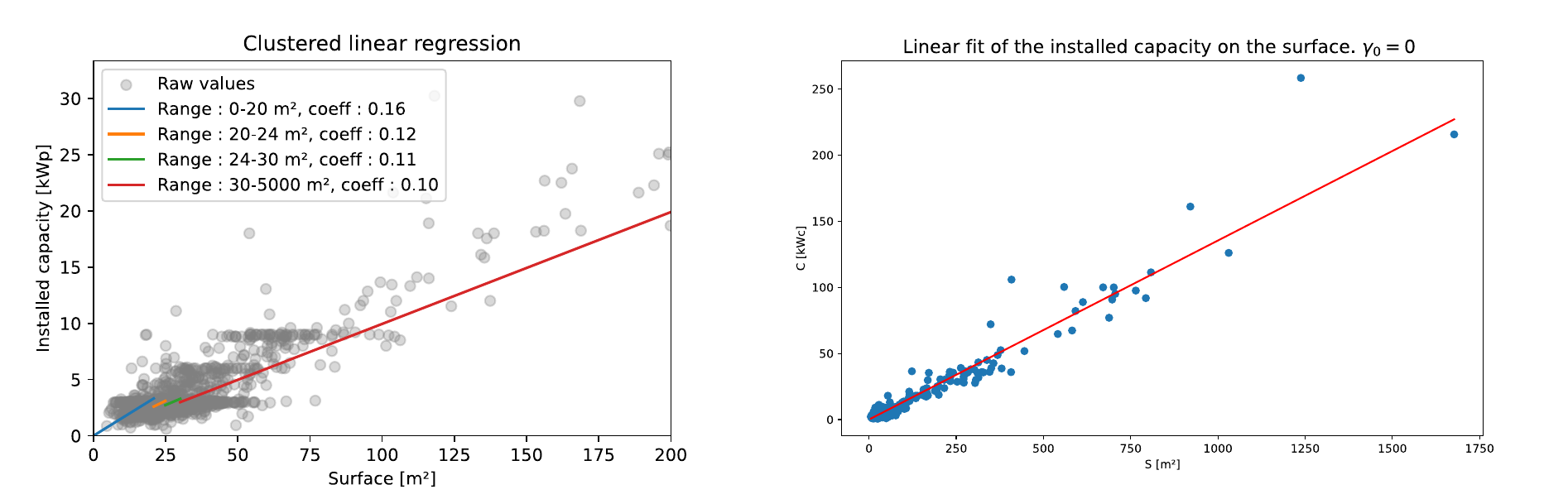}
    \caption{Left: clustered linear regression. Right: linear regression with a single coefficient.}
    \label{fig:linear-fit}
\end{figure}

Figure \ref{fig:linear-fit} represents the linear regression of the installed capacity on the surface and shows the relatively low dispersion of points around this mean. The leftmost plot shows the different coefficients depending on the surface cluster. We focus on surfaces lower than 200 m², where the density of installations is the highest. We can see that the efficiencies recorded in our reference registry are higher for smaller installations.

\subsection{Evaluation criteria of the methods}

We evaluate each method based on metrics and the execution time. The performance metrics are the following:

\begin{itemize}
    \item Mean error (bias) (ME): $\displaystyle{\frac{1}{n}\sum_{i=1}^n\hat{x}_i - x_i}$
    \item Mean absolute error (MAE): $\displaystyle{\frac{1}{n}\sum_{i=1}^n\vert \hat{x}_i - x_i\vert}$
    \item Root Mean Square Error (RMSE): $\displaystyle{\sqrt{\frac{1}{n}\sum_{i=1}^n(\hat{x}_i - x_i)^2}}$
    \item Mean absolute percentile error (MAPE): $\displaystyle{\frac{1}{n}\sum_{i=1}^n\frac{\vert \hat{x}_i - x_i\vert}{x_i}}$
\end{itemize}

We evaluate our methods on the BDAPPV dataset, introduced in further detail in section \ref{sec:data}. When training is required (e.g., random forests or lookup table), we consider an independent subset of the dataset so that the methods are all evaluated on unseen data samples. 

\section{Data}\label{sec:data}

\subsection{PV installations training database}

{\tt PyPVRoof} was developed on the PV characteristics dataset "BDAPPV" introduced by Kasmi et al. \cite{kasmi2023crowdsourced}. This dataset contains ground truth segmentation masks, images, and installation characteristics for more than 28000 installations in France and Western Europe, along with an explicit link between the segmentation masks and the installation characteristics for a subset of 8000 of them.

\subsection{Geographical information system (GIS) data}

We also rely on the surface models provided by the Institut Géographique National (IGN). These digital surface models (DSM) leverage two methods, the "photogrammetry" DSM and the LiDAR. Photogrammetry uses parallax to get the altitude points associated with each coordinate. Indeed, altitudes over an area are determined from different pictures from different points of view as nearer objects (from the aircraft carrying the aerial camera) move faster than distant objects. Such data is available almost everywhere in France, with a ground resolution of around 20 cm/pixel and an altimetric precision of around 150 cm. The photogrammetry DSM is accessible at IGN's Géoservices portal, accessible at this URL: {\tt https://geoservices.ign.fr/}. The second technique, called Light Detection and Ranging (LiDAR), calculates distances from the reflection of a light beam on a surface. This technique has an altimetric resolution of 10 cm/pixels. Even though LiDAR raw data is composed of point clouds with around 10 points/m2, we have decided to interpolate and rasterize it to a 20 cm/pixel resolution to use the same developed methods to infer tilts and azimuths. By the time this article was written, the LiDAR DSM did not cover all of France and is only accessible for demonstration purposes on the IGN's dedicated webpage, accessible at this URL: {\tt https://geoservices.ign.fr/lidarhd}.

\section{Results of the benchmark between methods}\label{sec:results}

The results provided in this section can be replicated using the data available on our public Zenodo repository \cite{tremenbert2023pypvroof}. {\color{red}
These results are preliminary results and we do not draw any conclusions from them as we need a better baseline for a fair comparison of our methods}. 

\subsection{Summarized results}

In this section, we report the accuracy results for each method and each characteristic. Our selected methods offer a snapshot of the best accuracy currently attainable. 

Table \ref{tab:results-aggregate} summarizes the accuracy results of these methods. We can see a large improvement gain when using LiDAR data, which we suspect can be even larger if the Theil-Sen algorithm is applied directly on the point cloud. We focused on rasters for a fair comparison with the photogrammetry DEM. Besides, a surprising result is that LiDAR data is better for azimuth angle than tilt estimation. An explanation for this is that azimut estimation is less sensitive to noisy data points in the ($z$) elevation direction than tilt estimation. 

We highly recommend using the Theil-Sen method if the DSM is precise enough (e.g., coming from LiDAR data). Otherwise, the bounding-box method is competitive at a much lower computational cost.

\begin{table}[h]
    \centering
    \begin{tabular}{|c|c|c|}
    \hline
       {\bf Characteristic}  &  {\bf Method} & {\bf Accuracy (RMSE) [unit]}\\
       \hline
       \hline
       Surface  & Direct computation & 6.86 [m²]\\
       \hline
       Tilt & LUT & 10.29 [°]\\
       \cline{2-3} 
       & Theil-Sen (LiDAR) & 14.69 [°]\\
       \hline
       Azimuth & Bounding-box & 32.76 [°]\\
       \cline{2-3}
       & Theil-Sen (LiDAR) & 4.38 [°]\\
       \hline
       Surface & Linear regression & 0.687 [kWp]\\
       \hline
    \end{tabular}
    \caption{Accuracy results of {\tt PyPVRoof}'s methods for PV panels characteristics extraction}
    \label{tab:results-aggregate}
\end{table}

\subsection{Detailed results}

\paragraph{Surface estimation}

Table \ref{tab:results-surface} reports the results. We observe small differences between annotated and predicted masks which assess the overall quality of the predicted masks. However, we give particular attention to the positive bias between masks and the surface reported in the characteristics file of BDAPPV, highlighting a tendency to overestimate the referenced surface area. Such bias is irrelevant since BDAPPV's surface area values cannot be assessed: for instance, such an overrepresentation of installations of 20 m² could result from a systematic roundup. 

\begin{table}[h]
    \centering
    \begin{tabular}{|c|c|c|c|c|}
    \hline
        {\bf Method}  & {\bf ME}  & {\bf MAE} & {\bf RMSE} & {\bf Runtime} \\
         & [m²] &  [m²] &  [m²]  & [sec]\\
        \hline
        \hline
        Direct computation & 3.62 & 5.01 & 6.86 & (-)\\
        \hline
    \end{tabular}
    \caption{Performance metrics for the estimation of the projected surface. The mean surface area is 20 m².}
    \label{tab:results-surface}
\end{table}

\paragraph{Tilt angle estimation}

Table \ref{tab:results-tilt} presents preliminary results. For tilt estimation, it turned out that the LUT was a surprisingly strong baseline over the other methods: the random forest yielded only minor improvements, but the runtime is an order of magnitude larger. Although not significant for a single installation, such a difference in runtime is significant when scaling the method to thousands of PV polygons. As for the methods that require surface models, we can see that their accuracy relies on the quality of the input data. We tested the Theil-Sen method on photogrammetry-based surface models and LiDAR surface models. We can see a noticeable improvement when shifting from photogrammetry to LiDAR. { \color{red} We cannot yet compare the LUT and the Theil-Sen methods. The results displayed in table \ref{tab:results-tilt} are only preliminary.}

\begin{table}[h]
    \centering
    \begin{tabular}{|c|c|c|c|c|}
    \hline
        {\bf Method}  & {\bf ME} & {\bf MAE} & {\bf RMSE} & {\bf Runtime} \\
        & [°] & [°] & [°] & [sec]\\
        \hline
        \hline
        Random Forest & 6e-4 & 5.34 & 7.03 & 0.28\\
        \hline
        Look-up table & -2.40 & 7.68 & 10.29 & 6e-6\\
        \hline
        Theil-Sen & 3.99 & 14.10 & 17.50 & 0.09 \\
        (Photogrammetry) & &  &  &\\
        \hline
        Theil-Sen & 2.06 & 11.08 & 14.69 & 0.09\\
        (LiDAR) & &  & & \\
        
        \hline
        Hough with DSM & 2.90 & 13.45 & 16.62 & 2.47\\
        \hline
    \end{tabular}
    \caption{Performance metrics for the estimation of the tilt angle. The two lines for the Theil-Sen method report the accuracy results whether photogrammetry DSM or LiDAR DSM are passed as inputs.}
    \label{tab:results-tilt}
\end{table}

\paragraph{Azimuth angle estimation}

For azimuth estimation, we replicated the method of \cite{edun2021unsupervised} using the Hough algorithm. They report MAEs ranging from 15.62 to 30.53 degrees depending on the type of panel considered, the largest errors being associated with rooftop panels and the smallest with ground panels. Our replication is, therefore, in line with theirs, as we report an MAE of 22.70 degrees. Surprisingly, we see very few improvements brought by the Hough method with surface models. On the other end, the bounding-box method, which solely relies on the PV polygon, is a very accurate approach, even outperforming the Theil-Sen algorithm (in the case of photogrammetry DSM). The Theil-Sen method with LiDAR data is the most accurate, as shown in table \ref{tab:results-azimuth}.

\begin{table}[h]
    \centering
    \begin{tabular}{|c|c|c|c|c|}
    \hline
        {\bf Method}  & {\bf ME} & {\bf MAE}  & {\bf RMSE} & {\bf Runtime}  \\
        & [°] & [°] & [°] & [sec]\\
        \hline
        \hline
        Hough\cite{edun2021unsupervised} & 2.10 & 22.70 & 40.26 & 0.04\\
        \hline
        Hough with DSM & -0.73 & 23.78 & 43.66 &2.50\\
        \hline
        Theil-Sen & -6.65 & 15.54 & 35.64 & 0.09\\
        (Photogrammetry) &  &  &  & \\
        \hline        
        Theil-Sen & -0.08 & 3.10 & 4.38 & 0.09\\
        (LiDAR) &  &  &  & \\
        \hline
        Bounding-box & -1.39 & 12.90 & 32.76 & 0.02\\
        \hline
    \end{tabular}
    \caption{Performance metrics for the estimation of the azimuth angle. The two lines for the Theil-Sen method report the accuracy results whether photogrammetry DSM or LiDAR DSM are passed as inputs.}
    \label{tab:results-azimuth}
\end{table}

\paragraph{Installed capacity estimation}

Estimating the installed capacity requires the tilt angle. Indeed, we use the real surface rather than the projected surface as input to estimate the installed capacity. Rausch et al. \cite{rausch2020enriched} reported a 9 percentage point increase in the median absolute percentage error (MedAPE) for estimating the installed capacity when considering the tilt angle. We compared variants of the random forest estimator, with $\theta$ coming from different methods, to see how potential errors propagated. As it can be seen from table \ref{tab:results-ic}, all random forests perform equally. We can also see that these methods are only slightly better than the clustered linear regression, which improves over \cite{so2017estimating}. \cite{so2017estimating} reported mean squared errors ranging from 1.64 to 1.69, corresponding to an RMSE of 1.28-1.30. We slightly improved over their baseline with our clustered linear regression approach. 

\begin{table}[h]
    \centering
    \begin{tabular}{|c|c|c|c|c|c|c|}
    \hline
        {\bf Method}  & $\theta$ & {\bf ME} & {\bf MAE} & {\bf RMSE} & {\bf MAPE}& {\bf Runtime}\\
        &&[kWp]& [kWp] & [kWp] & [\%] & [sec]\\
        \hline
        \hline
        Random forest    & RF & 0.022 & 0.328 & 0.750 & 9.37 & 1.1e-1\\
        (with $S_{est}$) &  & & & && \\
        \hline
        Random forest    & TS  & 0.061 & 0.393 & 0.848 &11.48 & 4.4e-4\\
        (with $S_{proj}$ and $\theta$) &  && & & & \\
        \cline{2-7}
        Random forest&RF & 0.079 & 0.379 & 0.921 & 10.66 & 4.3e-2\\
        (with $S_{proj}$ and $\theta$) &   && & & & \\
        \hline
        Clustered linear regression & RF & -0.015 & 0.376 & 0.687 & 11.57 & 7.2e-7\\
        \hline
    \end{tabular}
    \caption{Performance metrics for the estimation of the installed capacity. Column $\theta$ indicates the method used to derive the tilt necessary to compute the estimated surface $S_{est}$, taken as input to estimate the installed capacity.}
    \label{tab:results-ic}
\end{table}

\subsection{Choice of the methods}

Based on the results of table \ref{tab:results-aggregate} and the comprehensive comparison reported in section \ref{sec:results}, we restricted ourselves to the following methods for each characteristic:
\begin{itemize}
	\item \underline{Surface}: direct computation.
	\item \underline{Tilt angle}: constant imputation, a look-up table, and Theil-Sen estimation. The constant imputation works in all cases, the LUT turned out to be very competitive compared to the random forests, and Theil-Sen is competitive when surface models are available.
	\item \underline{Azimuth angle}: we keep the bounding-box method and the Theil-Sen estimation to be used when surface models are available. 
	\item \underline{Installed capacity}: we keep the constant imputation and the linear models, as it turned out to be very competitive with the random forests.
\end{itemize}

\section{Conclusion} \label{sec:conclusion}

This work describes a method for the comprehensive and automated extraction of PV systems characteristics. This method takes as input PV polygons and returns the set of associated characteristics using the current best methods available and taking as input as few auxiliary data sources as possible. The user can pick his preferred method depending on the auxiliary data available in his case. Overall, we designed our approach to cover mostl use cases encountered when characterizing PV installation. In the context of growing PV and the increasing availability of large-scale inventories containing PV polygons, such methods will be beneficial for integrating these data into official registries or power forecasting models. Moreover, by enabling better characterization of the rooftop PV fleet, this package can contribute to producing statistics on PV installations characteristics.






\bibliographystyle{elsarticle-num} 
 \bibliography{cas-refs}

\appendix

\section{Benchmarked characteristics extraction methods}\label{sec:complementaty-methods}

\subsection{Random forests}

A random forest is an ensemble learning method that can be used for classification or regression problems. The output of the random forest is the average of the prediction of each regression tree. Random forests perform better than individual trees and are less prone to overfitting. All random forest estimators are trained on the BDAPPV dataset \cite{kasmi2023crowdsourced}, following a standard train/test split procedure. 

\paragraph{Tilt estimation} For the tilt estimation, the input variables are the localization of the PV panel polygon and its projected surface. We also tested the variant with additional input variables such as the length and width of the PV polygon, with no effect on the final accuracy. The dependent variable of the random forest, in this case, is the tilt angle of the installation. 

\paragraph{Installed capacity estimation} For the installed capacity, the dependent variables are the localization of the PV panel and the real surface, which corresponds to the projected surface corrected by the cosine of the tilt angle as defined in equation (\ref{eq:surface}). We consider two variants for this estimator: one with the tilt and projected surface passed separately as input and another one where we directly input the real surface as input. As further discussed in table \ref{tab:results-ic}, the model with the real surface yields the best accuracy results.

\subsection{Hough algorithm} 

This method is based on \cite{edun2021unsupervised}.
We first apply a Canny Edge Detector \cite{canny1986computational} to the installation mask and dilate the output borders with a 5x5 pixels kernel. Such dilatation aims to simplify the line detection while applying the Hough Transform \cite{hough1959machine}. This transform detects straight lines in binary edge-detected images by counting the number of aligned points with each point of the image. A line is accounted for when this number is above some defined threshold. We then calculate the angle and the length of each individual line, but contrary to \cite{edun2021unsupervised}, we do not select the angle with the maximum frequency because such a choice could lead to a 90 degrees-error for installations that are taller than they are wide. Indeed, we prefer to store the cumulated length per 1-degree bin and select two angles following two rules. The first angle is associated with the maximum cumulated length. Then, the second angle is associated with the second cumulated length amongst angles that are at least 40 degrees away from the first one. This way, we avoid selecting the same main orientation twice with a 1-2-degree difference.

\begin{figure}[h]
    \centering
    \includegraphics[width=\textwidth]{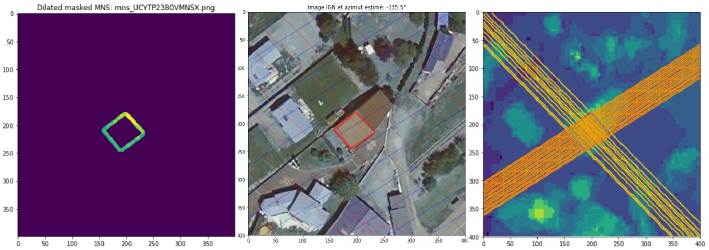}
    \caption{Estimation of the tilt angle using the Hough transform method}
    \label{fig:hough}
\end{figure}

We also improve upon \cite{edun2021unsupervised} by enhancing the method using the DSMs. DSMs can then be used to remove the ambiguity between the two main orientations given by the statistical analysis of cumulated lengths. We perform a quick test over the altitudes overlapping with the mask area to guess the approximate direction and value of the slope and select the closest angle given by the Hough method. Such a test, which can be a simple regression or the mean altitude difference along two opposite borders, allows estimating the tilt angle $\theta$ at the same time. Figure \ref{fig:hough} summarizes the process.

The leftmost image corresponds to the mask’s dilated edges with the corresponding altitude values. The Hough transform is applied to get the center image where every red segment corresponds to a detected line. Finally, the statistical analysis of lengths and angles is performed on the segments to get the two main orientations highlighted in the rightmost picture (the one corresponding to the highest drop in altitudes is in orange, and the remaining one is in yellow). 

\end{document}